\documentclass[twocolumn]{aastex701}
\usepackage{graphicx,times} 
\usepackage{natbib}
\usepackage{hyperref}
\usepackage{tcolorbox}
\usepackage{physics}
\usepackage{amssymb,amsmath}
\usepackage{indentfirst}
\usepackage{threeparttable}

\usepackage{graphicx} 
\usepackage{physics}
\usepackage{ragged2e}
\usepackage{rotating}
\usepackage{amsmath}
\usepackage[version=4]{mhchem}
\usepackage{threeparttable}

\definecolor{forestgreen}{rgb}{0.13, 0.55, 0.13}

\newcommand{\ssst}{\scriptscriptstyle}
\newcommand{\E}[1]{\times 10^{#1}}

\newcommand{\s}{\,{\rm s}}      
\newcommand{\ps}{\,{\rm s}^{-1}}
\newcommand{\yr}{\,{\rm yr}}    

\newcommand{\cm}{\,{\rm cm}}    
\newcommand{\km}{\,{\rm km}}
\newcommand{\kms}{$\km\ps$}

\newcommand{\pc}{\,{\rm pc}}
  
\newcommand{\K}{\,{\rm K}}

\newcommand{\Tmb}{T_{\rm mb}}
\newcommand{\Tk}{T_{\rm k}}

\newcommand{\nHH}{n({\rm H}_{2})} 
\newcommand{\NHH}{N({\rm H}_{2})}
\newcommand{\VLSR}{V_{\ssst\rm LSR}}

\newcommand{\twCO}{$^{12}$CO}   
\newcommand{\thCO}{$^{13}$CO}

\newcommand{\NtwCO}{N({\rm ^{12}CO})}   
\newcommand{\HH}{H$_2$}       

\newcommand{\Jotz}{$J$=1--0}    
\newcommand{\Jtto}{$J$=2--1}
\newcommand{\Jttt}{$J$=3--2}

\begin{document}

\title{JCMT \twCO~\Jttt\ observations of Tycho's supernova remnant: constraints on the environmental gas properties}

\correspondingauthor{Ping Zhou, Samar Safi-Harb}
\email{pingzhou@nju.edu.cn, samar.safi-harb@umanitoba.ca}

\author[0009-0009-5591-9042]{Sen-di Bo}
\affiliation{Department of Astronomy, Nanjing University, Nanjing 210023, People's Republic of China}
\email{211850101@smail.nju.edu.cn}

\author[0009-0007-5968-2236]{Yu Huang}
\affiliation{Purple Mountain Observatory, Chinese Academy of Sciences, 10 Yuanhua Road, Nanjing 210023, People's Republic of China}
\affiliation{School of Astronomy and Space Science, University of Science and Technology of China, Hefei 230026, People's Republic of China}
\email{yuhuang@pmo.ac.cn}

\author[0000-0002-5683-822X]{Ping Zhou}
\affiliation{Department of Astronomy, Nanjing University, Nanjing 210023, People's Republic of China}
\affiliation{Key Laboratory of Modern Astronomy and Astrophysics, Nanjing University, Ministry of Education, Nanjing, 210023, People's Republic of China}
\email{pingzhou@nju.edu.cn}

\author[0000-0002-9776-5610]{Tian-Yu Tu}
\affiliation{Department of Astronomy, Nanjing University, Nanjing 210023, People's Republic of China}
\email{tianyutu@smail.nju.edu.cn}

\author[0000-0001-6189-7665]{Samar Safi-Harb}
\affiliation{Department of Physics and Astronomy, University of Manitoba, Winnipeg, MB, R3T 2N2, Canada}
\email{Samar.Safi-Harb@umanitoba.ca}

\author[0000-0002-7299-2876]{Zhi-Yu Zhang}
\affiliation{Department of Astronomy, Nanjing University, Nanjing 210023, People's Republic of China}
\affiliation{Key Laboratory of Modern Astronomy and Astrophysics, Nanjing University, Ministry of Education, Nanjing, 210023, People's Republic of China}
\email{pmozhang@gmail.com}

\author[0000-0002-4753-2798]{Yang Chen}
\affiliation{Department of Astronomy, Nanjing University, Nanjing 210023, People's Republic of China}
\affiliation{Key Laboratory of Modern Astronomy and Astrophysics, Nanjing University, Ministry of Education, Nanjing, 210023, People's Republic of China}
\email{ygchen@nju.edu.cn}

\author[0000-0003-2062-5692]{Hidetoshi Sano}
\affiliation{Department of Physics, Nagoya University, Furo-cho, Chikusa-ku, Nagoya, Aichi 464-8601, Japan}
\affiliation{Faculty of Engineering, Gifu University, 1-1 Yanagido, Gifu, Gifu 501-1193, Japan}
\email{sano.hidetoshi.w4@f.gifu-u.ac.jp}

\begin{abstract}

Recent observations suggest that Tycho's supernova remnant (SNR; SN~1572) is expanding into a cavity wall of molecular clouds (MCs), which decelerate the SNR and influence its multi-wavelength morphology. 
To constrain the physical properties of environmental MCs and search for heated gas, we perform a JCMT \twCO~\Jttt~observation and compare with previous \twCO~\Jtto, \twCO~\Jotz\ and \thCO~\Jotz\ data.
We present the \twCO~\Jttt~map toward Tycho and show that the \twCO~\Jttt\ spatial distribution and line profiles are similar to those of the lower-$J$ CO lines.
By comparing the multiple transitions of CO and the RADEX (Radiative transfer code in non-Local Thermodynamic Equilibrium) models, we constrain the physical properties of molecular gas surrounding Tycho: the northern cloud has a molecular column density of $\NHH=0.5$ -- $4.5\times 10^{22}$ cm$^{-2}$, while other regions have $\NHH=0.2$ -- $3.9\times10^{21}$ cm$^{-2}$; the kinetic temperatures $\Tk$ of these clouds are in the range of 9 -- 22 K and the volume densities $\nHH$ are 20 -- $700 \cm^{-3}$. 
We also discuss the difficulty in finding hot molecular gas shocked by such a young SNR.
We estimate that the shocked molecular layer can be as thin as 0.003 pc, corresponding to $0.2''$ at the distance of 2.5 kpc, which is 2 orders of magnitude smaller than the angular resolution of current CO observations. Therefore, our molecular observations are largely insensitive to the thin shocked gas layer; instead, they detect the environmental gas.
\end{abstract}


\section{Introduction}
\label{sec:Introduction}

Supernova remnants (SNRs) are extended structures formed through the dynamic interaction between supernova materials and the surrounding medium. The ambient medium plays a crucial role in governing SNR evolution, influencing the SNR morphology from the radio up to the $\gamma$-ray regime \citep[see][and references therein]{vink20}.
Observations have revealed that many SNRs are evolving in the molecular clouds \citep[MCs; e.g.,][]{huang86,Frail96,zhou23}, the densest phase of the interstellar gas with typical gas density over $10^2~\cm^{-3}$.
The SNR shock can perturb, compress, and heat the molecular gas, leading to high excitation of molecules and even enhanced abundances of some molecular species \citep{Disheock1993,Seta98}. 
In return, the MCs can efficiently slow down the shock, increase the postshock density, and thus affect the SNR dynamical evolution and multi-wavelength radiation.

Molecular shells have been identified around core-collapse SNRs, where the powerful stellar winds from their massive progenitor stars likely carved out molecular cavities prior to the supernova explosions \citep{chevalier99,chen13}. These molecular structures preserve crucial information about the progenitor winds through their radii and expansion dynamics, thereby offering insights into the progenitor masses. This has led to the intriguing possibility that the dense gas shells observed near Type Ia SNRs might similarly encode information about their progenitor systems. 
In recent years, an increasing number of Type Ia SNRs have been suggested to interact with molecular shells \citep[Tycho, N103B, and 3C 397,][]{zhou16,chen17,sano18,Sano25} or a cloudy medium  \citep[SN1006,][]{miceli14, Sano22}.

Tycho's SNR (a.k.a. SN 1572, 3C 10, G120.1+1.4) originated from a Type Ia supernova explosion occurred in AD 1572 \citep{Ruiz04,Baade1945}. The distance of Tycho's SNR is obtained to be in the range of 1.5--$\sim 5$ kpc 
\citep[see][and references therein]{hayato10}, with a large uncertainty depending on the methods.
Tycho is morphologically surrounded by a molecular shell structure at a local standard of rest (LSR) velocity $\VLSR=-61~\km\ps$, which corresponds to a distance of $\sim 2.5$~kpc \citep{Lee04, Jinlong2011, zhou16}. 
The preshock gas density, as revealed by the X-ray and infrared observations, is 0.1--$0.2~\cm^{-3}$, with a greatly enhanced density in the northeast \citep{Katsuda10,Williams13}.
The radio and X-ray proper motion measurements found a smaller expansion velocity in the eastern and northeastern shell \citep{reynoso97, Williams16}. 
The asymmetries in shock velocities can be naturally interpreted if this remnant has encountered a denser medium in the northeast. 
The dense gas is suggested to be part of a molecular shell that expands with a velocity of $\sim 5~\km\ps$. This expanding shell favors a single-degenerate (SD; a white dwarf and a non-degenerate star) progenitor system over the double-degenerate (DD; two white dwarfs) for Tycho  \citep{zhou16} since the SD system can drive accretion outflow while the DD system does not \citetext{e.g. \citealt{Hachisu96, Li97, Hachisu99}}. This differs from an earlier suggestion of DD origin of Tycho through the plasma ionization history study \citep{Badenes07}. 
Recently, the new X-ray proper motion study found that Tycho has experienced faster deceleration in the past 15 yr regardless of direction, supporting that the SNR is not just impacting dense gas in the northeast but encountering a cavity wall \citep{Tanaka20}. This reconciles the large density contrast between the X-ray observation and molecular gas, and reinforces the SD origin.

The high-resolution molecular observations of Tycho remain very limited, although the molecular shell has been observed using several low-$J$ CO transitions \citep{Lee04, Jinlong2011, zhou16}. 
The physical properties of Tycho's environment, especially the molecular density and association with Tycho, require further constraints. 
On the one hand, the recent and future evolution of Tycho is highly influenced by its environmental density.  On the other hand, the density of the ambient clouds strongly influences the hadronic $\gamma$-ray flux as the cosmic-ray protons collide with the clouds \citep{zhang13,xing22,kobashi25}.

In this paper, we present \twCO~\Jttt\ maps toward Tycho's SNR using the James Clerk Maxwell Telescope (JCMT), with an angular resolution ($14''$) much higher than previous data of the same transition \citep{Jinlong2011}. The goal is to constrain the density and excitation conditions of molecular gas near the SNR shell by comparing the multiple transitions of CO molecules.

In Section \ref{sec:Observation}, we introduce our JCMT CO observation and other data used in this work. In Section \ref{sec:result}, we show the molecular gas distribution and line ratio maps of \twCO~\Jttt/\Jtto\ and \Jtto/\Jotz. 
In Section \ref{sec:disscussion}, we calculate the physical properties of molecular gas by comparing the observations and the radiative transfer model RADEX, and discuss the difficulty of finding heated gas and broadened profiles with molecular observations. The conclusion is provided in Section~\ref{sec:conclusion}.

\section{Observations and Data Reduction}
\label{sec:Observation}

\subsection{JCMT observations}
\label{subsec:JCMTobs}
The observations were performed  in August, 2021 with the 
JCMT , which is a 15 m single-dish radio telescope located at an altitude of 4,092 m in Mauna Kea, Hawaii.
We use the heterodyne array receiver program (HARP) on JCMT to observe the \twCO~\Jttt~emission line at the rest frequency of 345.796 GHz covering a $16' \times 16'$ region centered at ($00^h 25^m21^s, 64^{\circ}08^{\prime}35^{\prime \prime}$, J2000) (PI: Samar Safi-Harb; JCMT project ID: M21BP017). 
The position-switched observation mode is adopted. The half-power beamwidth (HPBW) of each receptor is approximately $14''$, and the main beam efficiency at 345 GHz is 0.64. The bandwidth is 250 MHz while the initial frequency resolution is $\sim 30$ kHz, which corresponds to 0.026 \kms.

To reduce the \twCO~\Jttt~data, we run pipeline ``oracdr'' with ``REDUCE$\_$SCIENCE$\_$NARROWLINE'' recipe in the software suite \href{http://starlink.eao.hawaii.edu/starlink}{STARLINK}\footnote{\url{http://starlink.eao.hawaii.edu/starlink}} \citep{Cavanagh08} and obtain the baseline-corrected data cube with a pixel size of $7''$. Then we use command ``sqorst'' to resample the velocity resolution to 0.2 km~s$^{-1}$, which is sufficient for resolving lines. The root mean square (RMS) noise level of the main beam temperature ($T_{\mathrm{mb}}$) is 0.55 K for each pixel.

\begin{figure*}[ht]
\begin{center}
\includegraphics[width=0.95\textwidth]{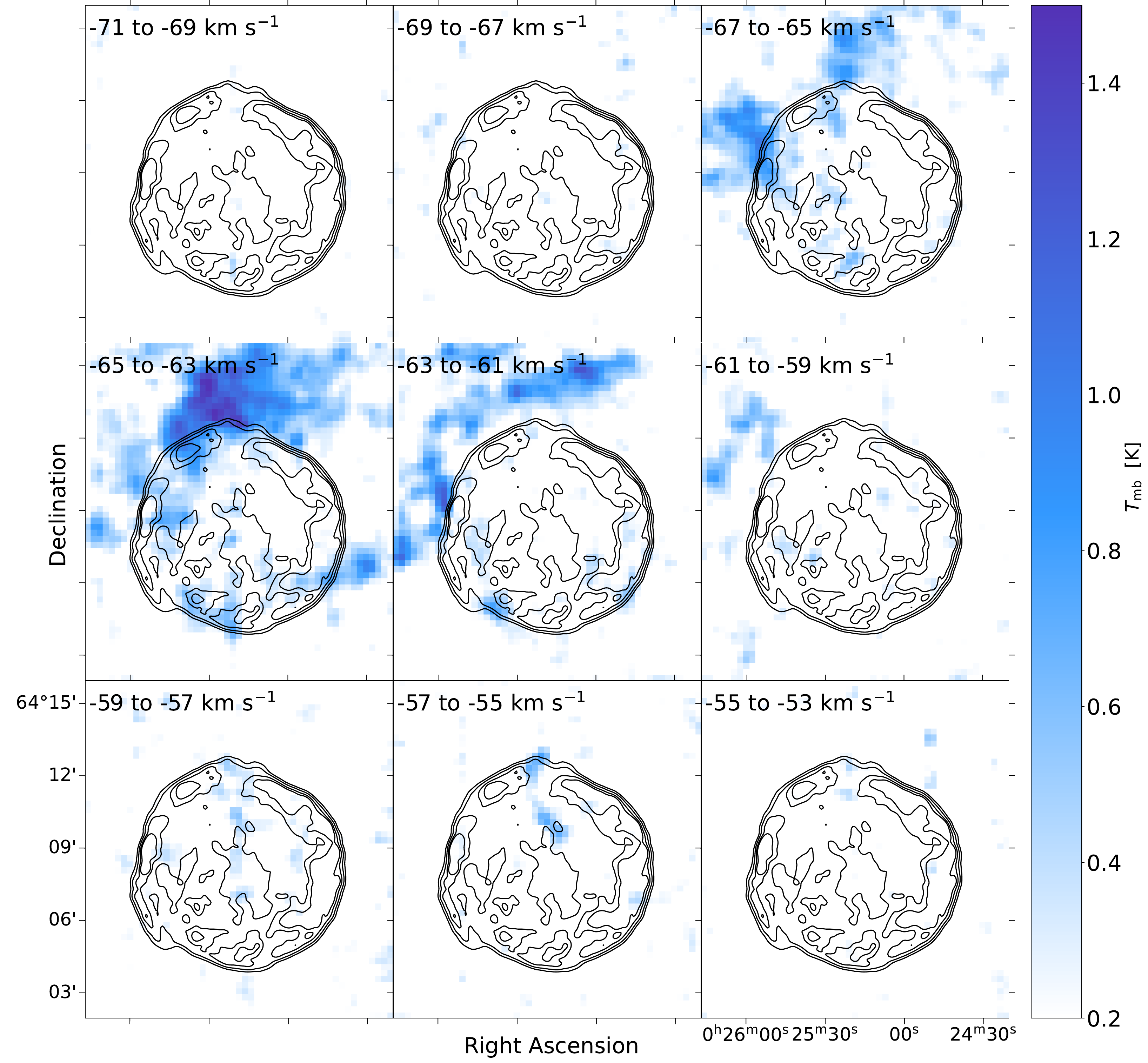}
\caption{The \twCO~\Jttt~velocity-averaged intensity maps with a step of 2 \kms\ ranging from $-71$ \kms\ to $-53$ \kms, overlaid with contours of the Chandra X-ray map. To increase the signal-to-noise ratio, the maps use the smoothed data with an angular resolution of 30$''$ and a pixel size of $15''$. Data with the intensity above $2\sigma$ are shown in the plot.}
\label{fig:co32map}
\end{center}
\end{figure*}

\begin{figure*}[h!]
\begin{center}
\includegraphics[width=0.95\textwidth]{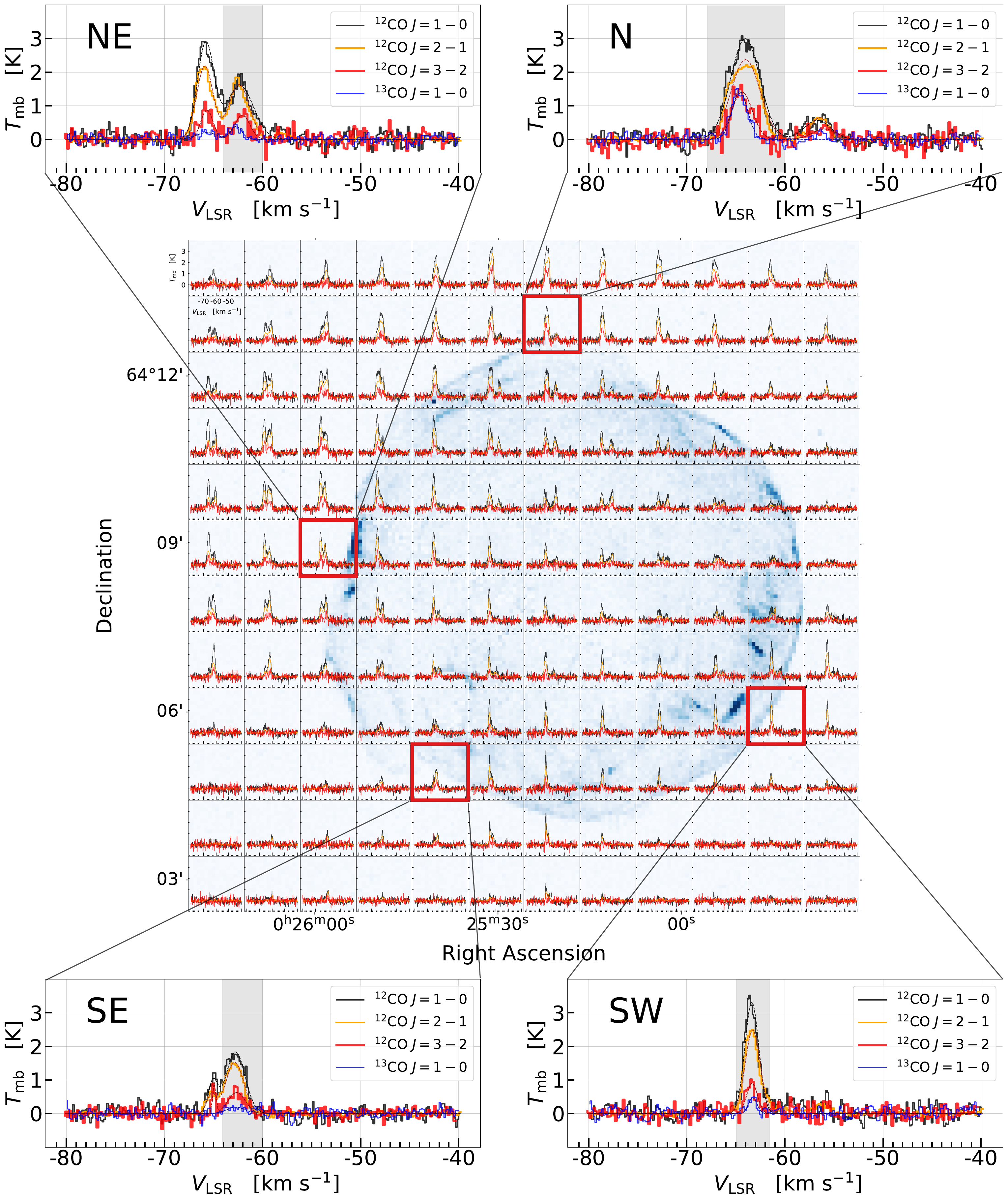}
\caption{The main panel shows a grid map of the line profiles of \twCO~\Jttt, \twCO~\Jtto~and \twCO~\Jotz~toward Tycho, overlaid with the Chandra hard X-ray flux in 4.1 -- 6.1~keV. The black, yellow and red lines represent \twCO~\Jotz, \twCO~\Jtto~and \twCO~\Jttt~line profiles, respectively. The zoomed-in panels illustrate spectra from selected regions ``N'', ``NE'', ``SE'', ``SW'', with \thCO~\Jotz~spectra (blue lines) added. The solid lines are the spectra we observed while the dashed lines represent fit results using the Gaussian function. The gray shadows denote the velocity components we use to study molecular gas properties with RADEX.}
\label{fig:line}
\end{center}
\end{figure*}

\subsection{Other data} 
\label{subsec:otherdata}

\subsubsection{IRAM 30m data}

The IRAM 30~m data \twCO~\Jtto~line emission cube is obtained from \citet{zhou16}. The original data cube has a velocity resolution of 0.2~\kms, a pixel size of $5.3''$, and an HPBW of $11.2''$. The mean RMS noise level of $T_{\mathrm{mb}}$ is 0.54 K in each pixel. 

\subsubsection{MWISP Data}

We retrieve the \twCO~\Jotz~and \thCO~\Jotz~ line emission cubes from the Milky Way Imaging Scroll Painting (MWISP) project \citep{su19}. The data have an angular resolution of $55''$, pixel size of $30''$, velocity resolution of 0.16 km s$^{-1}$, and mean RMS noise level of $T_{\mathrm{mb}}=0.38$ K in \twCO~\Jotz~and 0.067 K for \thCO~\Jotz. 

For comparison between the three CO transitions, the CO data cubes are further resampled to a uniform velocity resolution of 1 \kms. The final angular resolution and pixel size of three data cubes are smoothed and regridded to a beam of $55''$ and a pixel size of $30''$, matching those of the MWISP data. At this resolution, the RMS of \twCO~\Jttt, $J=2-1$ and $J=1-0$ are 0.095, 0.026 and 0.11 K, respectively. Since \twCO~\Jttt~and \twCO~\Jtto~observations have better angular resolution, we apply a smaller beam size of $30''$ and a pixel size of $15''$ when only comparing between the two transitions. 

\begin{figure*}[ht]
\begin{center}
\includegraphics[width=0.49\textwidth]{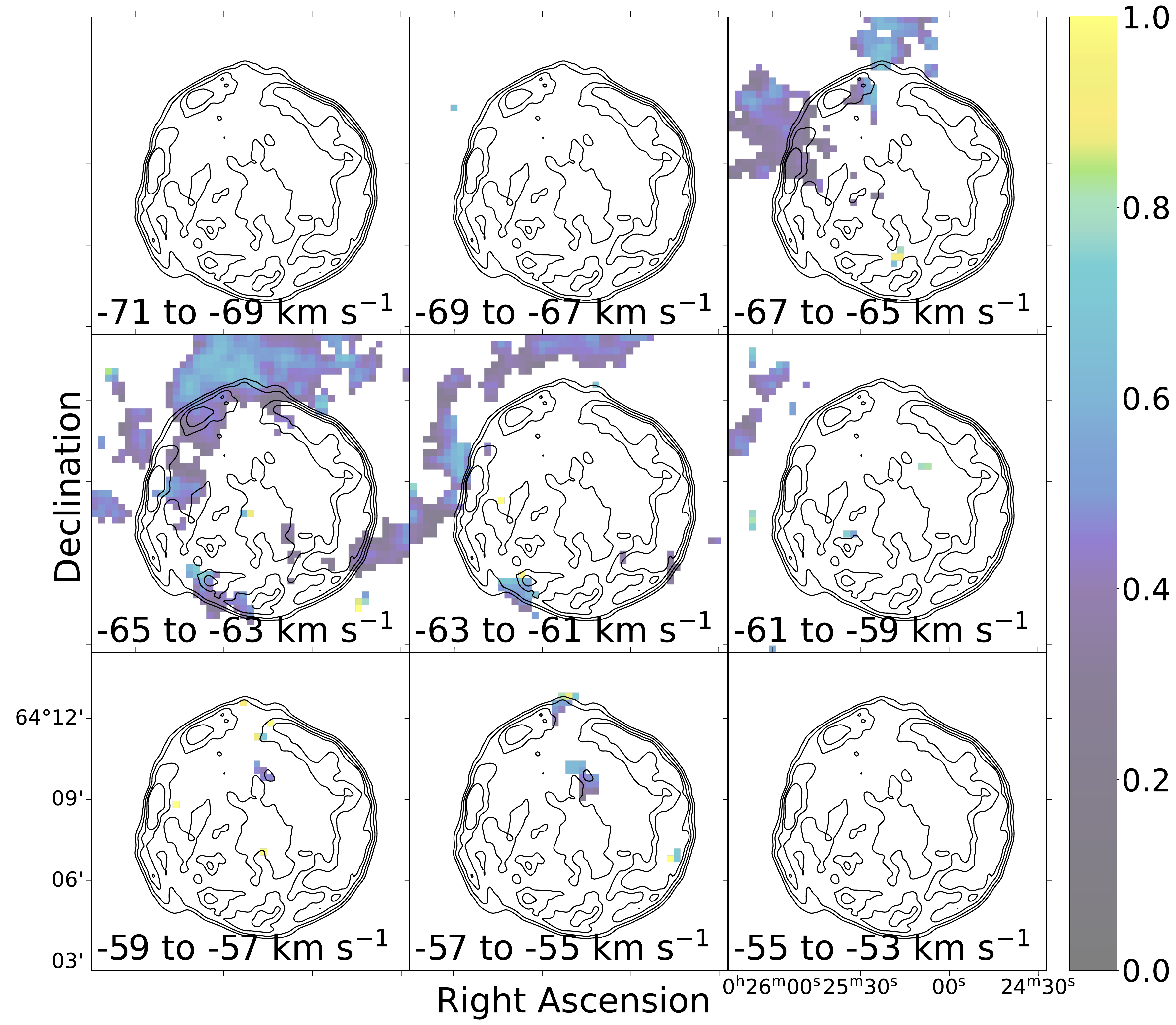}
\hfill
\includegraphics[width=0.49\textwidth]{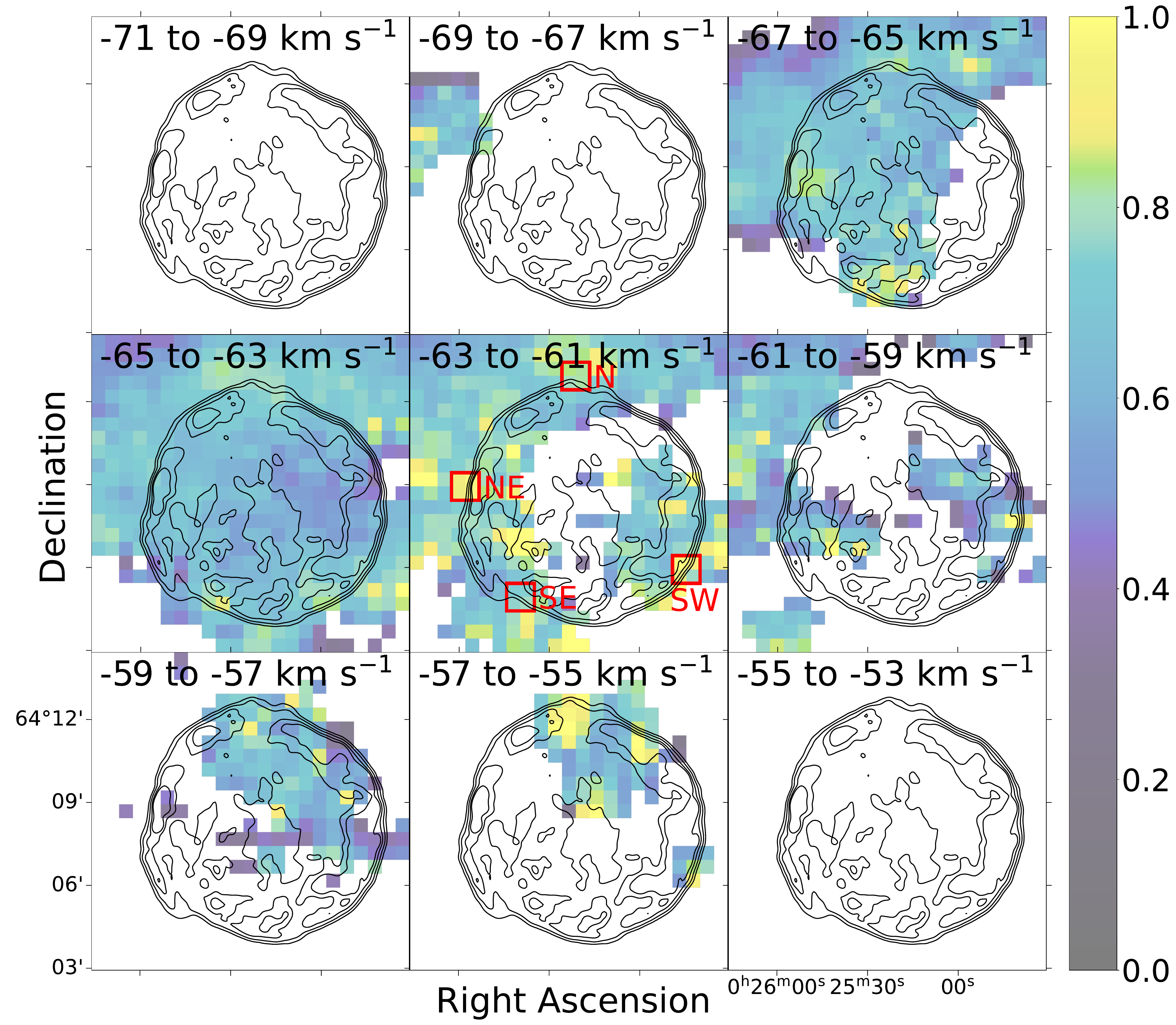}
\caption{The left panel displays the intensity ratio maps of \twCO~\Jttt~to \twCO~\Jtto\ from $\VLSR=-71~\km\ps$ to $\VLSR=-52~\km\ps$, while the right panel shows the ratio maps of \twCO~\Jtto~to \twCO~\Jotz. The spatial resolutions of the two maps are $30''$ and $55''$, respectively. The Chandra X-ray contours are overlaid. The red squares mark the four regions where we later calculate gas parameters using RADEX (see Table~\ref{tab:position}).}
\label{fig:ratio map}
\end{center}
\end{figure*}

\subsubsection{Chandra image}
To show the X-ray morphology and boundary of Tycho's SNR and to compare with the CO emission, we reduce a Chandra X-ray observation of Tycho's SNR observed in 2010 with a long exposure of 176~ks (obs.\ ID: 10095; PI: Hughes). We apply $fluximage$ in CIAO software (vers.\ 4.9) to obtain the X-ray image in the 4.1 -- 6.1~keV energy band, where the emission is dominated by synchrotron radiation.

\section{Results}
\label{sec:result}

Figure~\ref{fig:co32map} shows the velocity-averaged intensity maps of \twCO~\Jttt~emission toward Tycho's SNR in the LSR velocity ($\VLSR$) from $-71$ to $-53$ \kms.
We find relatively strong \twCO~\Jttt\ emission in the north and northeast in the velocity interval $\VLSR=- 67$ to $-61\km\ps$, with some faint emission in the south and southwest. 
The overall distribution of \twCO~\Jttt\ emission delineates the shell of the SNR, with a strip-like CO feature that aligns with the northwestern radio enhancement. This is consistent with previous CO observations \citep{Lee04, Jinlong2011, zhou16}. 

\setlength{\tabcolsep}{12pt}
\begin{table*}[t]
    \centering
    \caption{Line properties and the RADEX fit results for four regions (marked as red squares in Figure \ref{fig:ratio map}).}
    \scriptsize
    \begin{threeparttable}
    \begin{tabular}{c c c c c c c c}
        \hline\hline
         Position & Lines & $\VLSR$ & $ T_{\rm peak}$ & $dV$ & $T_{\rm k}$ & $\nHH$ & $\log(\NtwCO)$\tnote{*}  \\
         
         & & $\rm km~s^{-1}$ & K & $\rm km~s^{-1}$ & K & $\rm cm^{-3}$ & $\rm cm^{-2}$ \\
        \hline
        N & \twCO~\Jttt~ & $-64.36\pm0.08$ & 1.41$\pm$0.08 & 2.46$\pm$0.16 & ${10^{+1}_{-1}}$ & ${51^{+127}_{-34}}$ & ${18.25^{+0.51}_{-0.54}}$ \\
           & \twCO~\Jtto~& $-64.01\pm0.02$ & 2.37$\pm$0.03 & 3.48$\pm$0.05 & & & \\
           & \twCO~\Jotz~& $-64.04\pm0.03$ & 3.00$\pm$0.06 & 3.49$\pm$0.08 & & & \\
           & \thCO~\Jotz~& $-64.59\pm0.03$ & 1.42$\pm$0.05 & 1.90$\pm$0.08 & & & \\
        \hline
        NE & \twCO~\Jttt~& $-62.06\pm0.10$ & 0.72$\pm$0.08 & 1.81$\pm$0.23 & ${18^{+4}_{-3}}$ & ${48^{+103}_{-31}}$ & ${17.15^{+0.53}_{-0.43}}$ \\
           & \twCO~\Jtto~& $-62.39\pm0.02$ & 1.63$\pm$0.03 & 2.47$\pm$0.05 & & & \\
           & \twCO~\Jotz~& $-62.32\pm0.08$ & 1.60$\pm$0.06 & 3.29$\pm$0.21 & & & \\
           & \thCO~\Jotz~& $-62.67\pm0.10$ & 0.37$\pm$0.05 & 1.38$\pm$0.23 & & & \\
        \hline
        SE & \twCO~\Jttt~& $-62.64\pm0.08$ & 0.63$\pm$0.05 & 2.10$\pm$0.20 & ${13^{+4}_{-3}}$ & ${295^{+429}_{-193}}$ & ${16.64^{+0.44}_{-0.39}}$ \\
           & \twCO~\Jtto~& $-62.83\pm0.01$ & 1.52$\pm$0.02 & 2.18$\pm$0.03 & & & \\
           & \twCO~\Jotz~& $-62.76\pm0.05$ & 1.84$\pm$0.07 & 2.32$\pm$0.13 & & & \\
           & \thCO~\Jotz~& $-62.71\pm0.24$ & 0.20$\pm$0.04 & 2.17$\pm$0.60 & & & \\
        \hline
        SW & \twCO~\Jttt~& $-63.34\pm0.06$ & 0.90$\pm$0.08 & 1.32$\pm$0.14 & ${12^{+3}_{-3}}$ & ${98^{+291}_{-77}}$ & ${16.96^{+0.68}_{-0.45}}$ \\
           & \twCO~\Jtto~& $-63.28\pm0.01$ & 2.46$\pm$0.04 & 1.81$\pm$0.03 & & & \\
           & \twCO~\Jotz~& $-63.37\pm0.02$ & 3.33$\pm$0.10 & 1.64$\pm$0.06 & & & \\
           & \thCO~\Jotz~& $-63.27\pm0.08$ & 0.47$\pm$0.08 & 1.05$\pm$0.20 & & & \\
        \hline
    
    \end{tabular} 
    \begin{tablenotes}
    
    \footnotesize
    \item [*] $\NtwCO$ represents the observed column density within the $55''$ beam. This value is reproduced by the RADEX model as $\NtwCO_{\mathrm{obs}} = f \NtwCO_{\mathrm{RADEX}}$, where $f$ denotes the filling factor obtained from the same RADEX calculation.
    \end{tablenotes}     
    \end{threeparttable}
    \label{tab:position}
\end{table*}

Figure~\ref{fig:line} shows a grid map of the line profiles of \twCO~\Jttt, \twCO~\Jtto~and \twCO~\Jotz\ in the vicinity of Tycho. 
The major CO components are in $\VLSR$ ranging from $-68$ \kms~to $-60$ \kms, while there is a weak component in $\VLSR$ from $-58$ \kms~to $-54$ \kms. In some regions (mostly in the northeast), the major components consist of two emission components that peak around $\VLSR=-66$ \kms\ and $-61$ \kms, respectively. 
In almost all regions, the intensities of three spectral lines follow the trend $I_{^{12}\mathrm{CO}~J=1-0} > I_{^{12}\mathrm{CO}~J=2-1} > I_{^{12}\mathrm{CO}~J=3-2}$.
At the eastern boundary of Tycho (RA = 26h15m0.15s, DEC = $64^{\circ}8'41''$, J2000; region ``NE''), however, the intensity ratio \twCO~\Jtto/\twCO~\Jotz\ is elevated to $\sim 1$. The location of this region is consistent with where the radio rim is deformed and the shock is decelerated \citep{reynoso97, Katsuda10}.

The zoomed-in panels of Figure~\ref{fig:line} present detailed spectra from region NE'' and three other regions (N'', SE'', and ``SW'') selected at different azimuth angles along the rim for comparison. 
We use a single or a double Gaussian function to fit the line profiles (see zoomed-in panels in Figure \ref{fig:line}) and summarize the line properties of the velocity center, peak main-beam temperature $T_{\rm peak}$ and line width $dV$ (FWHM) in Table~\ref{tab:position}. We notice a possible offset between line centroids in region NE and N.

We have confirmed the morphological evidence of the association between Tycho and molecular gas. 
To strengthen the evidence of interaction, one needs to find some other evidence such as line broadening and gas heating if the shocked gas layer can be resolved \citep{Seta98, jiang10, Chen14}. 
The high \twCO~\Jttt/\Jtto\ ($R_{3-2/2-1}$) or \twCO~\Jtto/\Jotz\ ($R_{2-1/1-0}$) intensity ratios ($> 1$) can also be understood as the MC being optically thin and heated by the shock \citep{Sakamoto94}. 
However, the largest line width (full width at half maximum, FWHM) of MCs near Tycho is $\sim$ 5 \kms, which is not significantly larger than the line widths of unshocked ambient clouds in the northeast. 
Thus, the spectra do not exhibit the broadened lines or very high $R_{3-2/2-1}$, $R_{2-1/1-0}$ ($\gg 1$) ratios. 
This suggests that our molecular observations cannot resolve the thin molecular layer shocked by Tycho or there is no direct interaction between Tycho and MCs. In Section~\ref{sec:discussion2}, we will elaborate our interpretation and show that the expected thickness of the shocked molecular layer is two orders of magnitude smaller than the resolution of JCMT.

To examine the distribution of molecular properties near Tycho and further search for heated gas, we also plot the intensity ratio maps of \twCO~\Jttt~to \Jtto\ and \twCO~\Jtto~to \Jotz\ (see Figure~\ref{fig:ratio map}). Most high-ratio points are detected in $\VLSR -63$ to $-61$ \kms. 
At region ``NE", the value of $R_{2-1/1-0}$ reaches 1.1 at $\VLSR= -62.5$ \kms.

\section{Discussion}\label{sec:disscussion}

\subsection{Physical Properties of the Molecular Clouds }

\begin{figure*}[h]
\begin{center}
\includegraphics[width=0.49\textwidth]{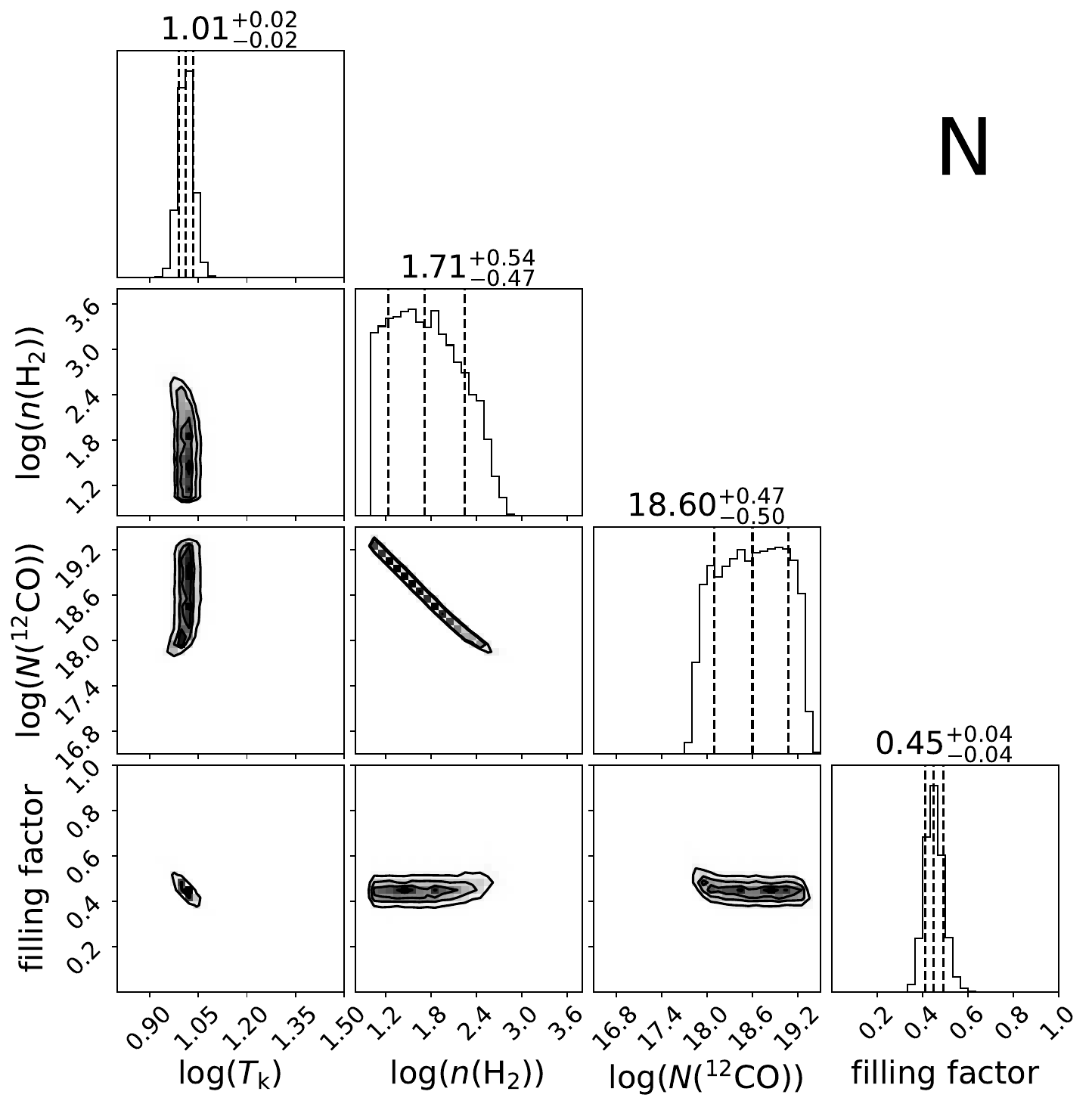}
\includegraphics[width=0.49\textwidth]{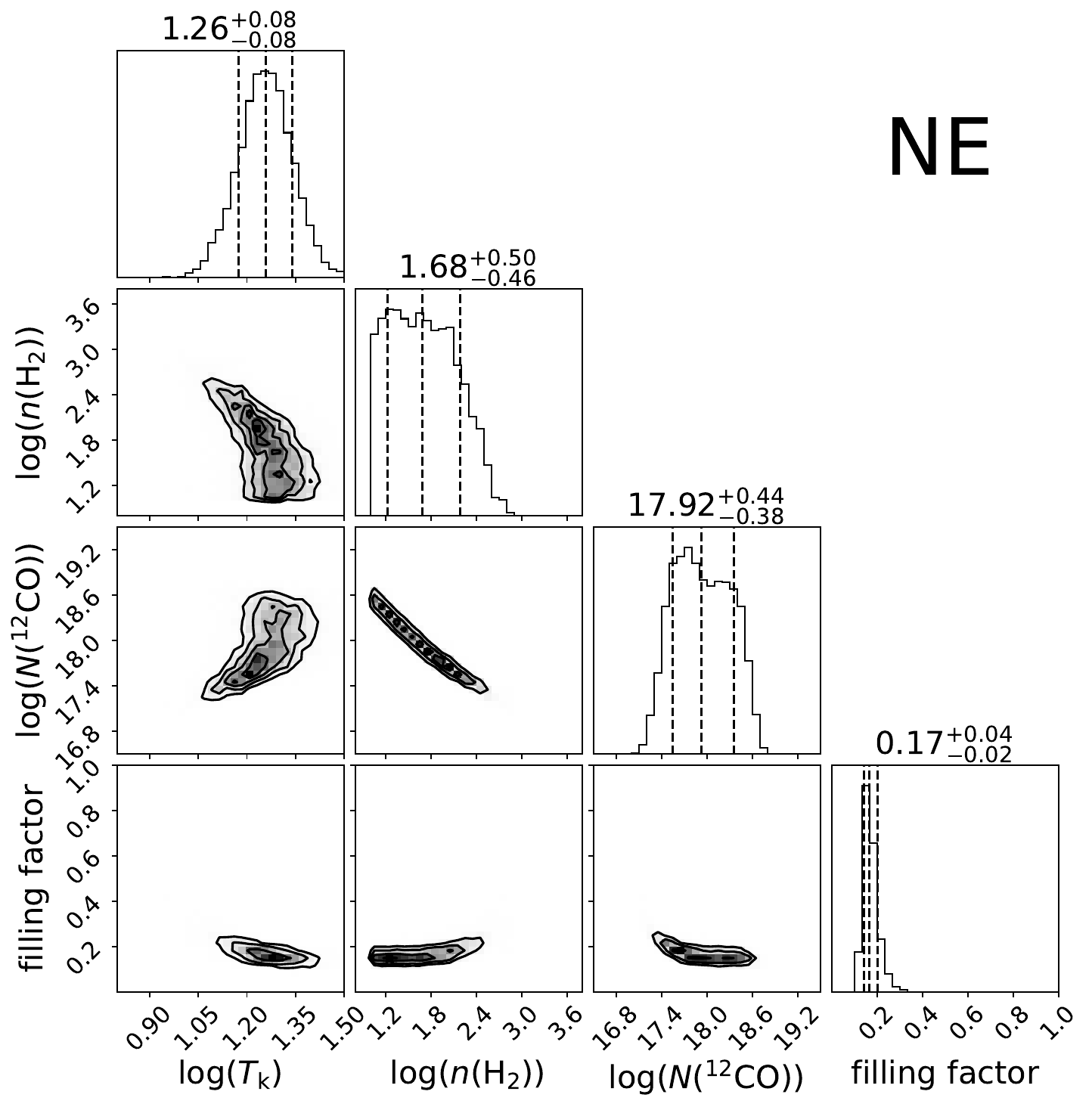}\\
\includegraphics[width=0.49\textwidth]{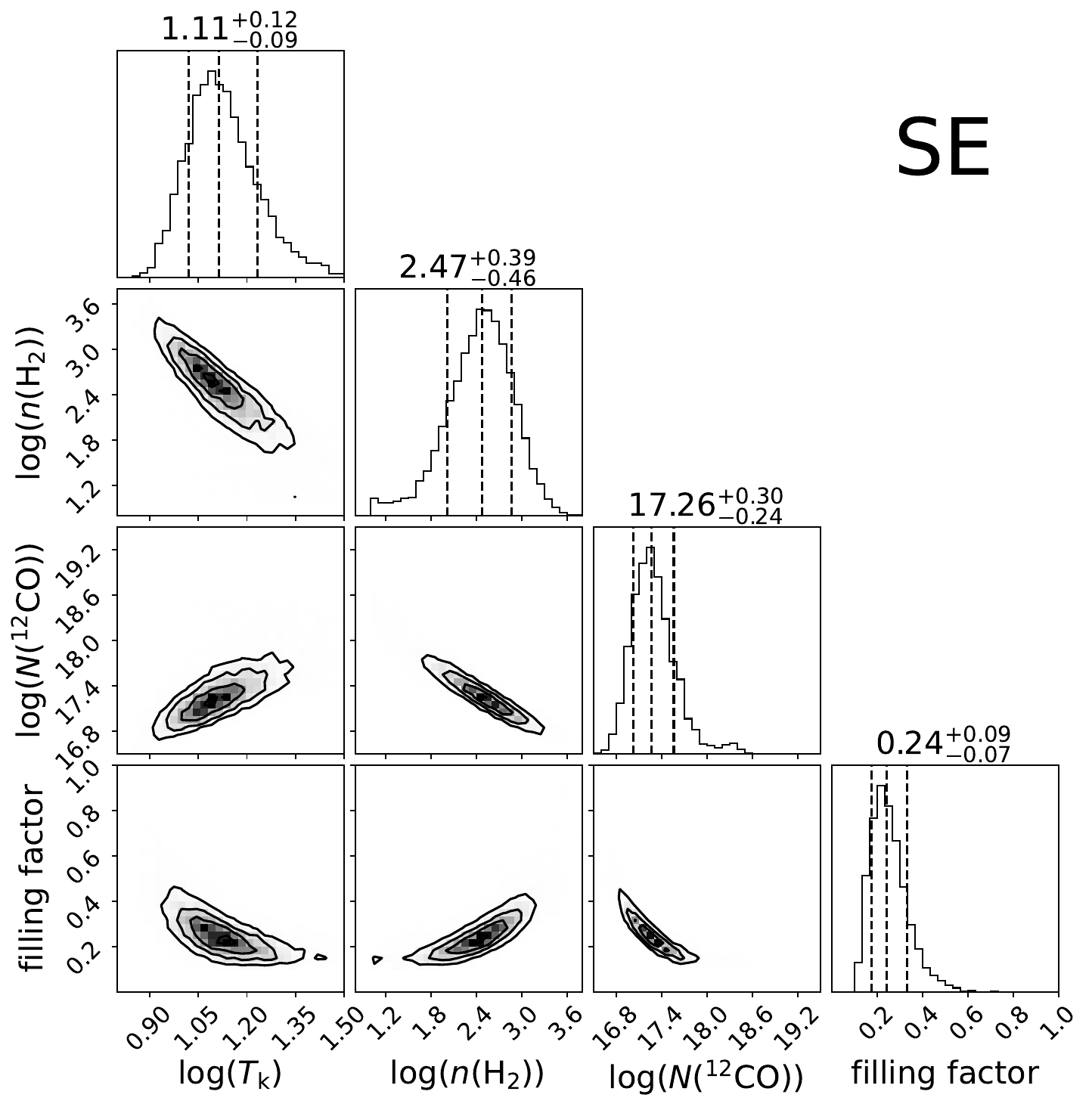}
\includegraphics[width=0.49\textwidth]{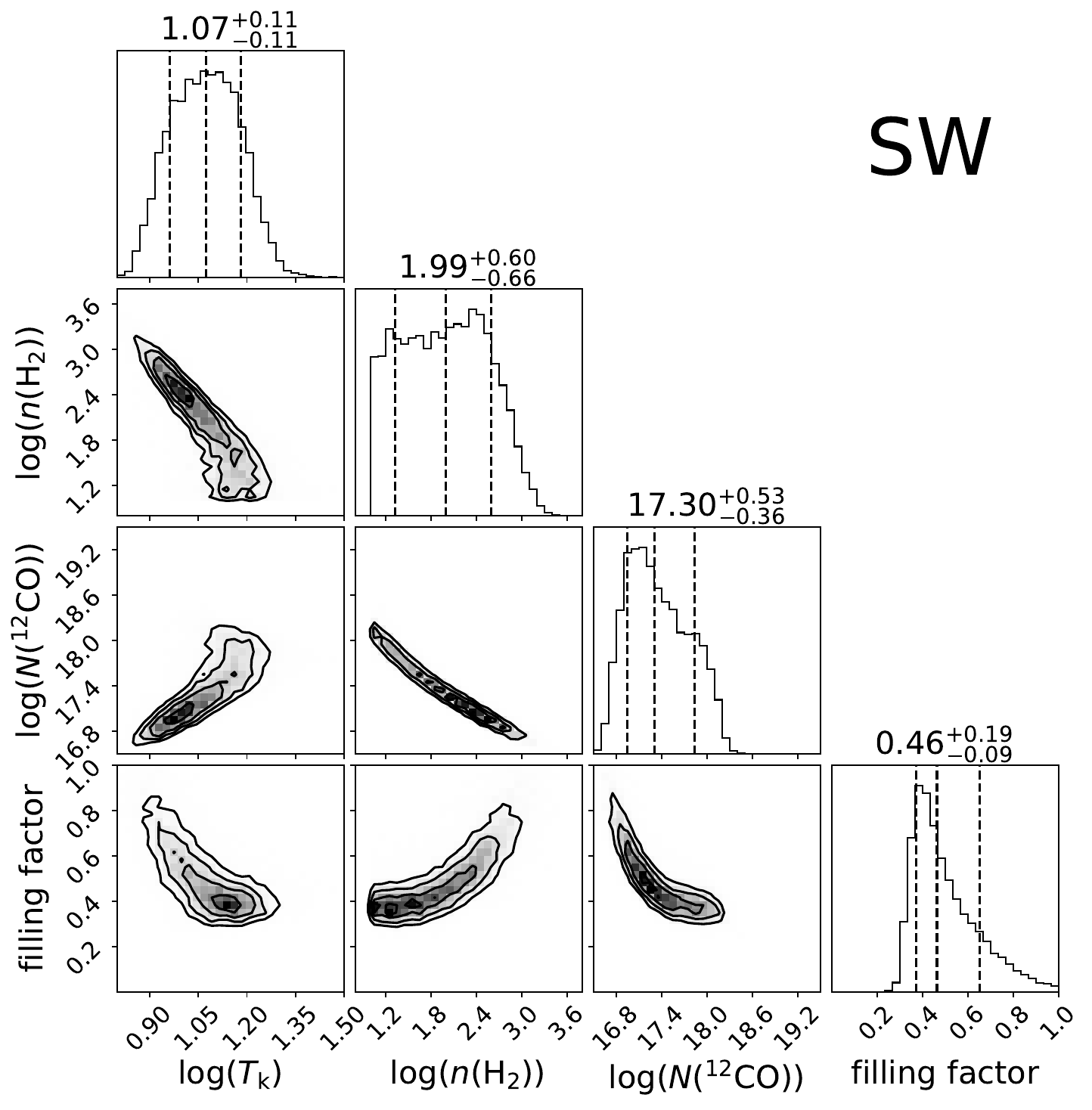}
\caption{MCMC corner plots based on multiple transitions of CO data and RADEX models. The dashed lines indicate the 0.16, 0.5, 0.84 quantiles of 1D PDFs.}
\label{fig:mcmc}
\end{center}
\end{figure*}

We calculate gas properties in four regions (see zoomed-in panels in Figure~\ref{fig:line} and red squares in Figure~\ref{fig:ratio map}).
The observed line properties of these regions are shown in Table \ref{tab:position}.
We used RADEX \citep[radiative transfer code in non-Local thermodynamic equilibrium;][]{van07} and the observed multi-$J$ CO emission to examine the physical properties of MCs toward Tycho. RADEX is a tool to calculate the non-LTE radiative transfer of specific physical conditions by using the escape probability formulation and assuming an isothermal and homogeneous medium. We calculate the brightness temperatures $T_{\rm B}$ of different CO transitions under various gas conditions including kinetic temperature $\Tk$, H$_2$ gas density $\nHH$, column density $\NHH$, and line width $dV$. The abundance ratios of 65 and $5\times 10^5$ are adopted for \twCO-to-\thCO\ and \HH-to-\thCO, respectively \citetext{\citealt{Dickman78, Vladilo91, Stahl92, Hawkins93, Langer93}}.

We compare the peak temperature ratios \twCO~\Jttt/\twCO~\Jtto, \twCO~\Jtto/\twCO~\Jotz, and \twCO~\Jtto/\thCO~\Jotz\ and the peak temperature of the \thCO~\Jotz\ emission with those predicted by RADEX modeling. We adopt the velocity width of the \thCO\ line for the RADEX models. For the escape probability method in RADEX, we consider the model for a homogeneous slab with a constant velocity gradient (SLAB). 
Using a different geometry, such as a homogeneous sphere, provides consistent results within a 1-$\sigma$ range. 
Monte Carlo Markov Chain (MCMC) method is used to provide the probability distribution for gas physical parameters including kinetic temperature $\Tk$, column density $N$(\twCO) and volume density $n$(\HH). 
The details of the MCMC method, such as assumptions, priors and burn-in length, are summarized in Appendix \ref{appendix}.

The corner maps for the MCMC results are shown in Figure~\ref{fig:mcmc} and the fit results are tabulated in Table~\ref{tab:position}. 
The \twCO\ column densities are constrained to be 0.5 -- $5.8\E{18}~\cm^{-2}$ in the north and 0.2 -- $4.8\E{17}~\cm^{-2}$ in other regions. 
Thus, considering the H$_2$/\twCO\ abundance ratio of $7700$, the $N$(\HH) is derived to be $0.4$ -- $4.5\times 10^{22}$ cm$^{-2}$ in the north and $0.2$ -- $3.9\times10^{21}$ cm$^{-2}$ in other regions.
At region ``NE'', we obtain $\NHH= 1.1^{+2.6}_{-0.7}\times 10^{21}~\cm^{-2}$, similar to that derived by \citet[$1.6 \times 10^{21}$ cm$^{-2}$]{zhou16}. 
The kinetic temperatures of the four regions are $10\pm 1$~K, $18^{+4}_{-3}$~K, $13^{+4}_{-3}$~K, and $12^{+3}_{-3}$~K, respectively, which are consistent with unheated molecular gas. 
The volume densities $\nHH$ are 20 -- $700 \cm^{-3}$ in these four regions. 
The present analysis should be interpreted with the caveat that it assumes single-component molecular clouds, since the observations lack the resolution or sensitivity to constrain more complicated multi-zone structures.

\subsection{On the lack of MC-shock interaction evidence} \label{sec:discussion2}

As aforementioned, our JCMT \twCO~\Jttt\ observation and the comparison with the IRAM 30~m \twCO~\Jtto\ and MWISP \twCO~\Jotz\ data do not provide further evidence about the physical interaction between Tycho and the nearby MCs, but also cannot refute the association.
The association between Tycho and dense ISM has been supported by both the proper motion studies \citep{reynoso97,Katsuda10, Williams17} and expansion of the molecular bubbles \citep{zhou16}. 
Here we discuss several reasons and implications about the lack of strong or broadened \twCO~\Jttt\ emission near Tycho.

Firstly, the shock velocity of Tycho reaches around 4000~\kms\ \citep{Williams17, Tanaka20} that can efficiently dissociate the molecules. For a shock propagating into a dense cloud, the cloud shock velocity $v_{\rm c}$ can be estimated by assuming a pressure equilibrium between the blast wave and the cloud shock: $n_c v_c^2 \sim  n_0 v_b^2$ \citep{McKee75}, where $n_{\rm c}$ and $n_0$  are the cloud and intercloud densities, respectively, and $v_b$ is the blast wave speed. Providing an averaged ambient density of $n_0\sim 0.1 \cm^{-3}$ \citep{Williams13} and $n_c\sim 100 \cm^{-3}$, we derived a cloud velocity $v_{\rm c} \sim 130 (n_{\rm c}/100)^{-0.5}~\km\ps$, which far exceeds the velocity threshold that can lead to molecule dissociation (25--$50~\km\ps$; \citealt{McKee80}). Therefore, broadened CO emission is expected to be ultra-weak or entirely absent, depending on the fraction of molecules that survive after the shock. 

Secondly, the shocked molecular layer could be too thin to be discerned with current JCMT observation ($15''$). The radio and X-ray measurements show the shock deceleration occurs only in the past two decades at the northeastern shell. The thickness of the shocked layer can be estimated as $l_c\sim v_c t_c=0.003~\pc~(v_{\rm c}/130~\km\ps) (t_c/20~\yr)^{-1}$, corresponding to $0.2''$ at the distance of 2.5~kpc. This thickness is two orders of magnitude smaller than the angular resolution of current CO observations. The beam of observation is mainly filled by emission from unshocked MCs and thus the line profiles do not reflect the heated/shocked gas. Given the short interaction time, the shock energy in Tycho has not been substantially transferred to the surrounding molecular gas. Another example is the young ($\sim$ 1800 yr) SNR RCW 86, which shows only a modest enhancement of the \twCO~\Jtto/\Jotz~ratio at the cloud surface \citep{Sano17}. In contrast, with long-term shock-cloud interaction, older SNRs such as W44, IC 443, and W28 show high \twCO~\Jtto/\Jotz~ratios and clear line broadening \citep[e.g.,][]{Seta98,reach05}. 

Observations with higher angular resolution are crucial to identify shocked gas in Tycho. Current observations have not resolved detailed molecular structures toward Tycho, as indicated by small beam filling factors (0.2--0.5) obtained from the MCMC analysis. For instance, with a beam size of $21''$, the IRAM \twCO~\Jotz\ lines reveal  $\Tmb\sim  2$--3~K, while the MWISP data with a beam size of $55''$ yield a lower value of about 1.6 K. We interpret that the severe beam dilution may explain the different line strengths observed between our studies. 
Moreover, \cite{zhou16} reported an elevated \twCO~\Jtto /\Jotz\ of 1.6 in the northeastern MC using the IRAM~30m \twCO~\Jtto\ data cube (beam size $=11''$) and \twCO~\Jotz\ single-position observation (beam size=$21''$).  Applying the MWISP \twCO~\Jotz\ data cube, we provided the \twCO~\Jtto/\Jotz\ ratio distribution across Tycho's SNR and found the ratio is mostly less than 1.1. We suggest that the larger beam incorporates more unheated gas with a lower line ratio, thereby diluting the contribution from the shocked layer and reducing the observed ratio. 
Therefore, further sub-arcsec high-resolution observations, such as IRAM/NOEMA, are needed to discern the shocked molecular layer and search for broadened lines. Finding such evidence will establish the association between Tycho and the expanding molecular bubble, and such support the SD scenario for the progenitor star. 
Furthermore, fast shock can dissociate and ionize molecular gas. \cite{arias19} reported the enhanced low-frequency absorption in the northeastern shell of Tycho, resulting from the dense gas layer ionized by either shock or ionizing radiation. Future radio observations with improved angular resolution in the low-frequency bands may constrain the thickness of the ionized layer and determine its morphology consistency with a shocked layer.

\section{Summary}\label{sec:conclusion}
We have performed a JCMT \twCO~\Jttt\ observation toward the Tycho's SNR and compared it with other CO observations to investigate the molecular gas properties. The results are summarized as follows:
\begin{enumerate}
  \item We present the \twCO~\Jttt\ maps towards Tycho. From  $\VLSR=-71$ to $-53$ \kms, the emission mainly distributes in north and northeast, with some faint structure in southeast and southwest. The intensities follow the trend $I_{^{12}\mathrm{CO}~J=1-0} > I_{^{12}\mathrm{CO}~J=2-1} > I_{^{12}\mathrm{CO}~J=3-2}$ except region ``NE'' with $I_{^{12}\mathrm{CO}~J=2-1}/I_{^{12}\mathrm{CO}~J=1-0}$ elevated to $\sim1$. We have not found very broad lines with the JCMT observations.
  \item 
  By comparing the four \twCO\ and \thCO\ transitions and the prediction from RADEX models, we estimate the physical properties of MCs toward Tycho. 
  The kinetic temperatures of these clouds are in the range of 9 -- $22$~K, consistent with unheated gas.
  The northern cloud has a molecular column density of $\NHH=0.4$ -- $4.5\times 10^{22}$ cm$^{-2}$ while other regions have $\NHH=0.2$ -- $3.9\times10^{21}$ cm$^{-2}$. The volume densities $\nHH$ are constrained to be 20 -- 700 $\cm^{-3}$.
  \item We discuss several reasons why we cannot identify the shocked gas. We derive the thickness of the shocked molecular layer of $0.2''$, which is two orders of magnitude smaller than the angular resolution of current CO observations. In this case, it is difficult to identify shock features in the clouds, since the beam is mainly filled with unshocked clouds. Higher-resolution observations are needed to discern the shocked molecular layer and search for broadened lines.
\end{enumerate} 

\begin{acknowledgements}
P.Z.\ acknowledges support from the National Natural Science Foundation of China  No.\ 12273010 and the Fundamental Research Funds for the Central Universities No.\ KG202502.
S.S.H. acknowledges support from the Natural Sciences Research Council of Canada through the Canada Research Chairs and the Discovery Grants programs.
 Z.-Y.Z. acknowledges the support of the National Natural Science Foundation of China (NSFC) under grants No. 12041305, 12173016, 1257030642, and 12533003.
 Y.C.\ acknowledges support from the NSFC grants 12173018, 12121003, 12393852, and 12573047.
 This work was also supported by JSPS KAKENHI grant Nos. 21H01136 (HS) and 24H00246 (HS).

 \twCO~\Jttt~observation was obtained by the James Clerk Maxwell Telescope, operated by the East Asian Observatory on behalf of The National Astronomical Observatory of Japan; Academia Sinica Institute of Astronomy and Astrophysics; the Korea Astronomy and Space Science Institute; the National Astronomical Research Institute of Thailand; Center for Astronomical Mega-Science (as well as the National Key R\&D Program of China with No. 2017YFA0402700). Additional funding support is provided by the Science and Technology Facilities Council of the United Kingdom and participating universities and organizations in the United Kingdom and Canada. 
This research made use of the data from the Milky Way Imaging Scroll Painting (MWISP) project, which is a multi-line survey in \twCO/\thCO/$\rm C^{18}O$ along the northern galactic plane with PMO-13.7m telescope. We are grateful to all the members of the MWISP working group, particularly the staff members at PMO-13.7m telescope, for their long-term support. 
MWISP was sponsored by National Key R\&D Program of China with grant 2017YFA0402701 and CAS Key Research Program of Frontier Sciences with grant QYZDJ-SSW-SLH047.

\end{acknowledgements}

\bibliography{references} 

\renewcommand{\thefigure}{A.\arabic{figure}}
\setcounter{figure}{0}

\appendix
\section{RADEX parameters and MCMC calculation}\label{appendix}

\begin{figure}[ht]
\centering
\includegraphics[width=0.47\textwidth]{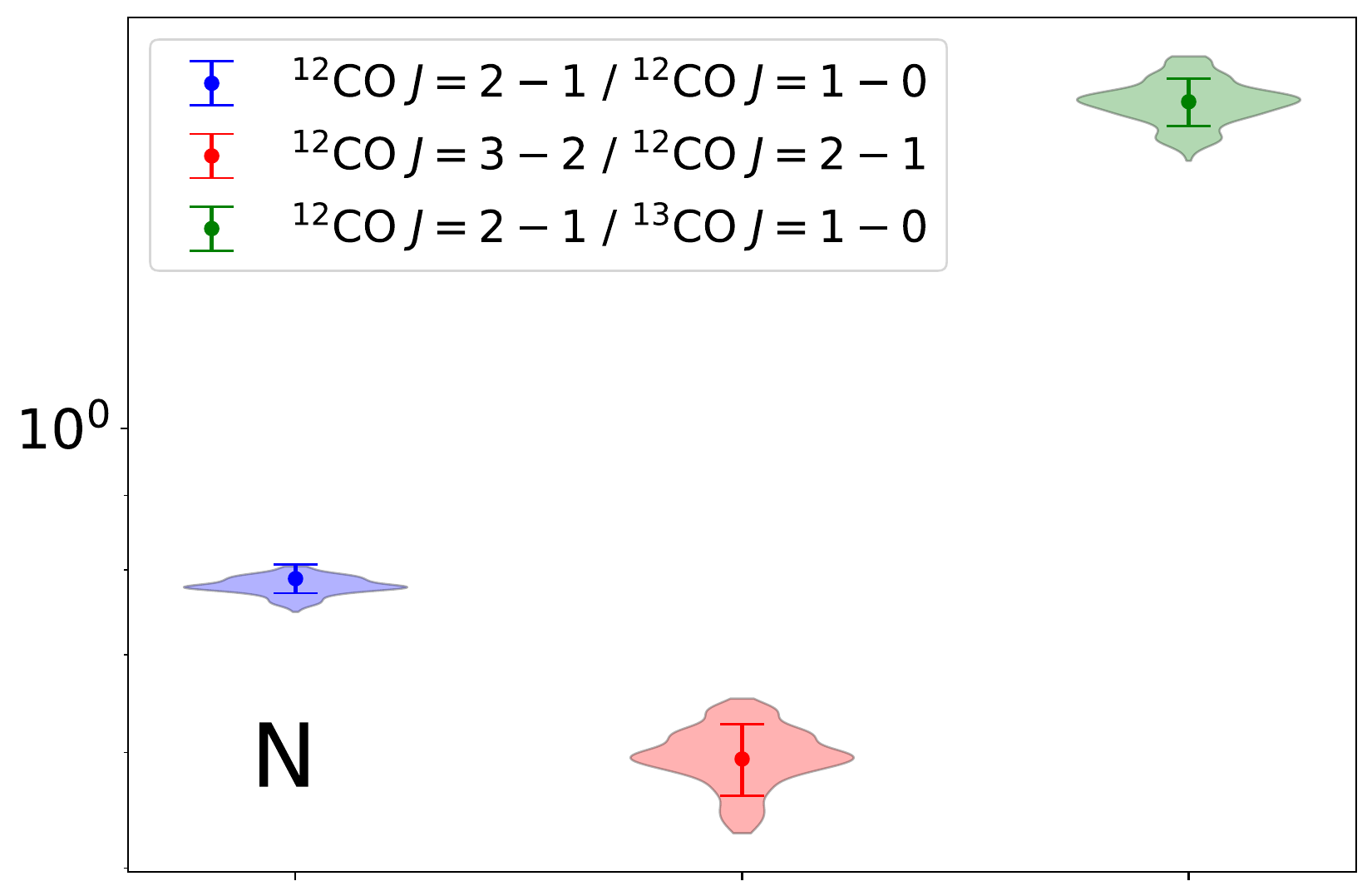}
\includegraphics[width=0.47\textwidth]{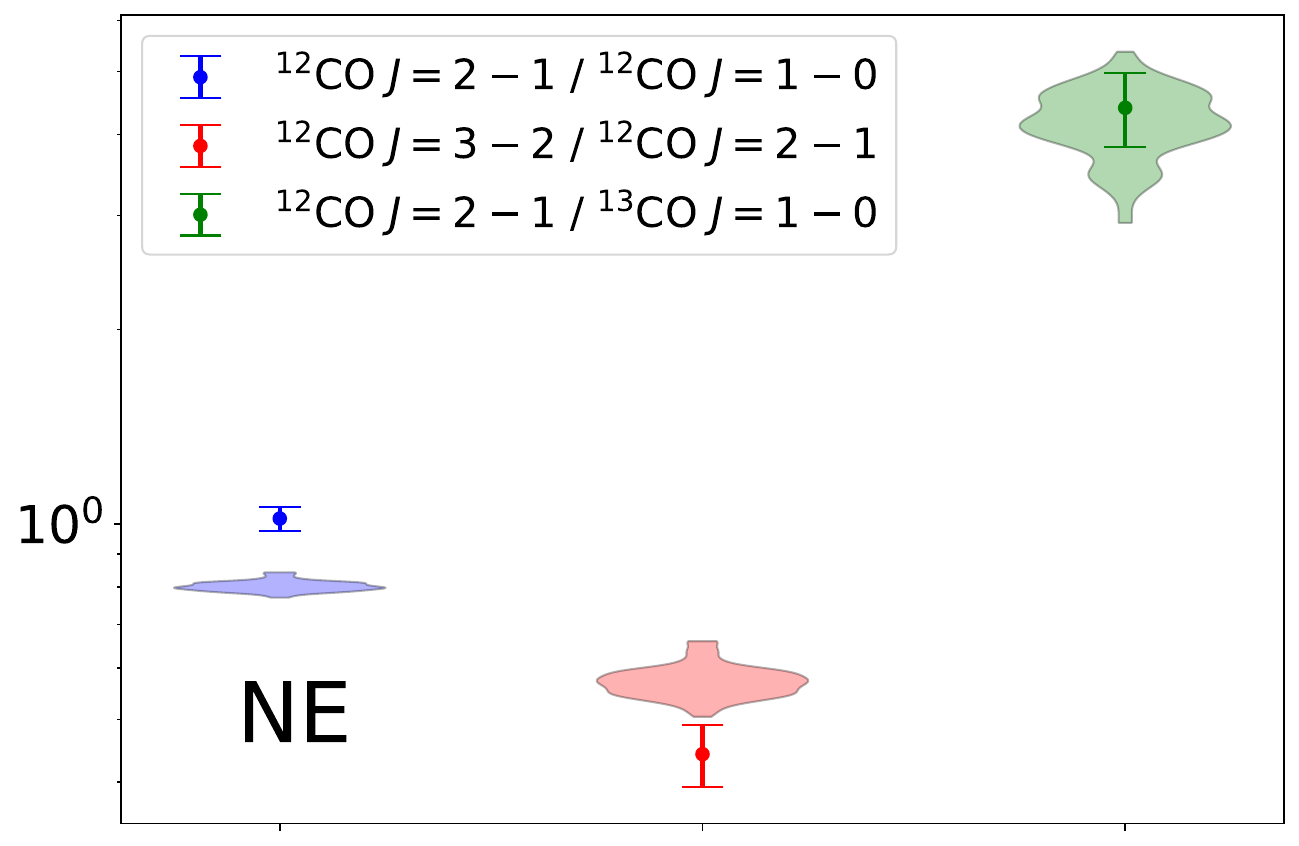}\\
\includegraphics[width=0.47\textwidth]{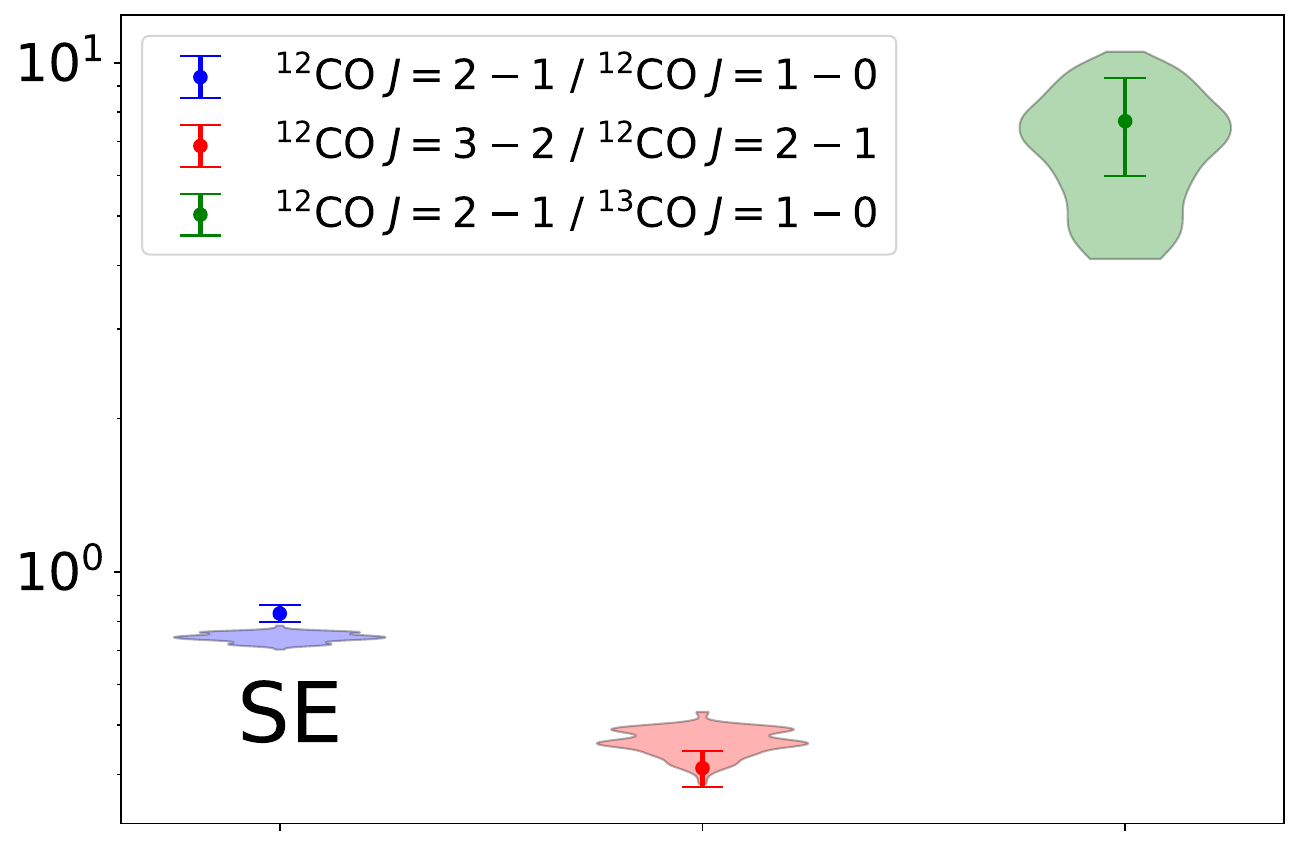}
\includegraphics[width=0.47\textwidth]{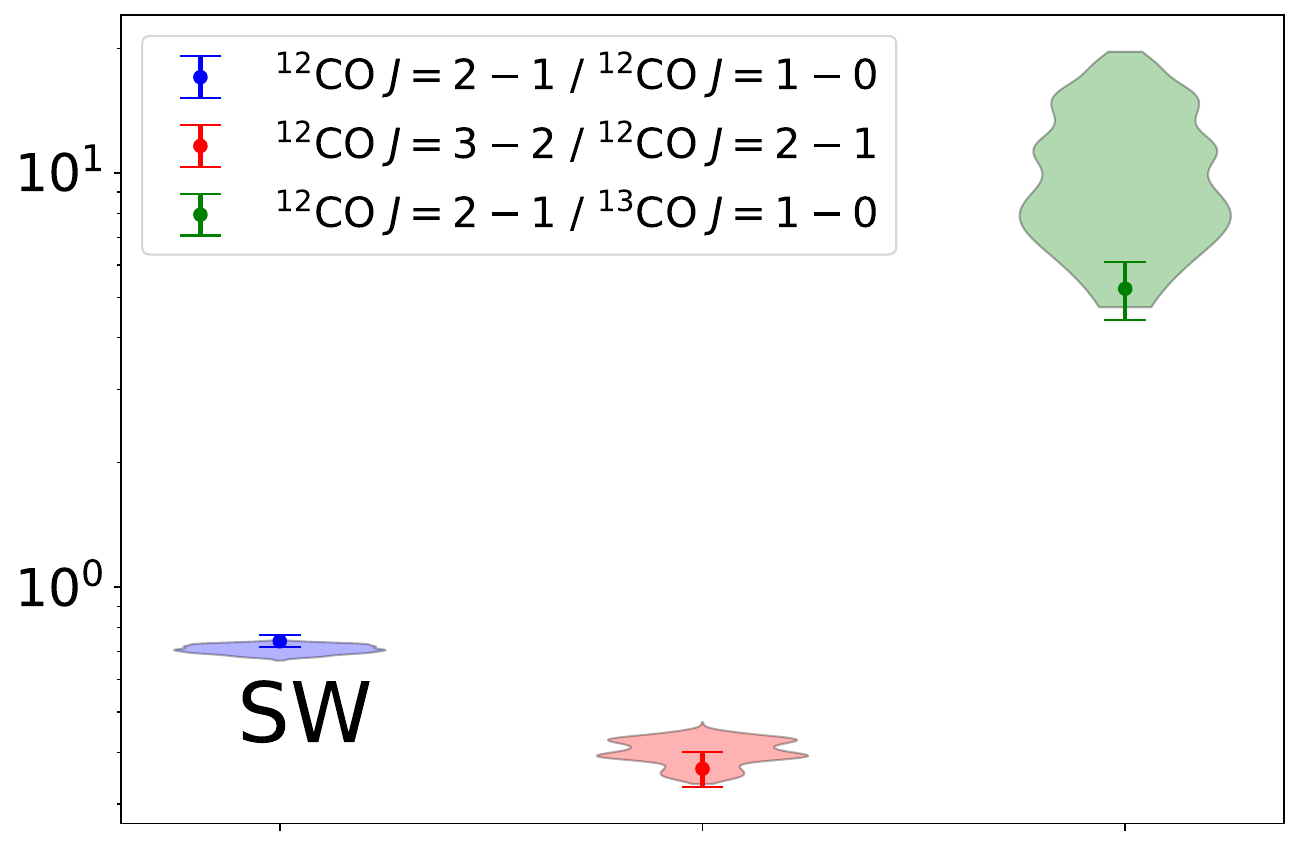}
\caption{The comparison between posterior distribution of line ratios in selected regions and the observed ones. The violin plots show the posterior distribution values from RADEX. The errorbars show the observed values.}
\label{fig:violin}
\end{figure}

We use molecular data files from the Leiden Atomic and Molecular Database (LAMDA) as the input of spectroscopic and collisional data in RADEX and assume that CO only collides with \HH. 
The parameter sampling space is logarithmically uniformly distributed, the same as the distribution of the random initial inputs.
Priors of $\Tk$, $\NtwCO$ and $\nHH$ are 10 -- 1000~K, $10^{11}$ -- $10^{22.5}\ {\rm cm^{-2}}$, $10$ -- $10^7 {\rm cm^{-3}}$, respectively. 
We adopt 10000 iterations for MCMC in each region, which includes 5000 burn-in steps. We run RADEX with the last 600 samples of physical parameters to get the posterior distribution of the intensity ratios to evaluate whether the fitting results are reliable by comparing with the observed values. 
The posterior distributions shown in Figure \ref{fig:violin} are similar to the observed values.

\end{document}